\documentclass[twocolumn,secnumarabic,showpacs,amssymb, amsmath,tightenlines,aps,prb]{revtex4}
\usepackage{longtable}
\usepackage{bm}
\usepackage{epsfig}
\usepackage{graphicx}

\begin{document}

\title[Ferromagnetism and Distortions in YTiO$_3$] {Ferromagnetism and Lattice Distortions in the
Perovskite YTiO$_3$}
\author{W. Knafo$^{1,2,3}$, C.
Meingast$^1$, and H. v. L\"{o}hneysen$^{1,2}$}

\address{$^1$Forschungszentrum Karlsruhe, Institut f\"{u}r Festk\"{o}rperphysik, D-76021 Karlsruhe, Germany\\
$^2$Physikalisches Institut, Universit\"{a}t Karlsruhe, D-76128
Karlsruhe, Germany\\
$^3$Laboratoire National des Champs Magn\'{e}tiques Puls\'{e}s,
UMR CNRS-UPS-INSA 5147, 143 Avenue de Rangueil, 31400 Toulouse,
Cedex 4, France}

\author{A. V. Boris$^{4,5}$, P. Popovich$^4$, N. N. Kovaleva$^{4,5}$,
P. Yordanov$^4$, A. Maljuk$^{4,6}$, R. K. Kremer$^4$, and B.
Keimer$^4$}

\address{$^4$Max-Planck-Institut f\"{u}r Festk\"{o}rperforschung, Heisenbergstra\ss e 1, 70569 Stuttgart, Germany.\\
$^5$Department of Physics, Loughborough University,
Leicestershire, LE11 3TU, United Kingdom.\\
$^6$Hahn-Meitner-Institut, Glienicker Str. 100, 14109 Berlin,
Germany. \vspace{0mm}}

\date{\today}

\begin{abstract}

The thermodynamic properties of the ferromagnetic perovskite
YTiO$_3$ are investigated by thermal expansion, magnetostriction,
specific heat, and magnetization measurements. The low-temperature
spin-wave contribution to the specific heat, as well as an Arrott
plot of the magnetization in the vicinity of the Curie temperature
$T_C\simeq27$ K, are consistent with a three-dimensional
Heisenberg model of ferromagnetism. However, a magnetic
contribution to the thermal expansion persists well above $T_C$,
which contrasts with typical three-dimensional Heisenberg
ferromagnets, as shown by a comparison with the corresponding
model system EuS. The pressure dependences of $T_C$ and of the
spontaneous moment $M_s$ are extracted using thermodynamic
relationships. They indicate that ferromagnetism is strengthened
by uniaxial pressures $\mathbf{p}\parallel \mathbf{a}$ and is
weakened by uniaxial pressures $\mathbf{p}\parallel
\mathbf{b},\mathbf{c}$ and hydrostatic pressure. Our results show
that the distortion along the $a$- and $b$-axes is further
increased by the magnetic transition, confirming that
ferromagnetism is favored by a large GdFeO$_3$-type distortion.
The $c$-axis results however do not fit into this simple picture,
which may be explained by an additional magnetoelastic effect,
possibly related to a Jahn-Teller distortion.

\end{abstract}

\pacs{75.30.-m,75.50.Dd,75.50.Ee,75.80.+q}

\maketitle

\section{Introduction}

ABO$_3$ perovskites exhibit a large variety of electronic and
magnetic properties \cite{goodenough04}. The titanate family
ATiO$_3$ recently attracted particular interest, since YTiO$_3$
orders ferromagnetically below the Curie temperature $T_C\simeq27$
K, whereas LaTiO$_3$ orders antiferromagnetically below the N\'eel
temperature $T_N\simeq150$ K
\cite{goral82,okimoto95,goodenough04,mochizuki04,pavarini05,komarek07}.
In these systems, the $S=1/2$ spins localized on the $Ti^{3+}$
ions are responsible for the magnetic properties. A change from
ferromagnetism to antiferromagnetism can be continuously tuned by
varying the lanthanum concentration $x$ in the alloys
Y$_{1-x}$La$_x$TiO$_3$, or by changing the lanthanide A (A =
Yb$\rightarrow$La) in the undoped ATiO$_3$
\cite{goral82,okimoto95,goodenough04,mochizuki04}. A
GdFeO$_3$-type distortion is driven by ion-size mismatch and
comprises rotations of the TiO$_6$ octahedra. It is responsible
for the distorted structure of the ATiO$_3$ crytals, with the
space group $Pbnm$. This distortion is more pronounced in YTiO$_3$
than in LaTiO$_3$, being favored by smaller A$^{3+}$ ions (A =
Y,La) \cite{mochizuki04,pavarini05}. In YTiO$_3$, an additional
elongation, by about 3 $\%$, of the TiO$_6$ octahedra is observed.
This distortion has been ascribed to staggered ordering of the
$t_{2g}$ orbitals (Ti$^{3+}$ ions)
\cite{akimitsu01,iga04,komarek07}. The switch from
antiferromagnetism to ferromagnetism in the ATiO$_3$ perovskites
is probably controlled by the extreme sensitivity of the magnetic
superexchange interactions to the distortions of the lattice
\cite{mochizuki04,pavarini05,solovyev06}. However, the mechanism
driving this transition is still a matter of considerable debate
\cite{pavarini05,komarek07,akimitsu01,iga04,zhou05,khaliullin03,craco06,okatov05,solovyev06}.
For a proper description of the magnetic properties, it is thus
crucial to carefully consider their dependence on the lattice
distortion.

In this article, we present a study of the thermodynamic
properties of YTiO$_3$. Experimental details will be given in
Section \ref{exp}. In Section \ref{thermodyn}, the specific heat,
thermal expansion, magnetization, and magnetostriction data
measured with magnetic fields applied along the easy $c$-axis will
be presented. In Section \ref{analysis}, these results will be
discussed and compared to the behavior expected within a
three-dimensional (3D) Heisenberg ferromagnetic model
\cite{ulrich02}. As a specific example, we will show data on the
typical 3D Heisenberg system EuS \cite{maletta82,maletta89}. In
Section \ref{discussion}, the relation between the distortion and
the magnetic properties will be discussed in the light of our
results. The dependence of the distortion on the A$^{3+}$ ionic
sizes, on uniaxial pressures, and on the temperature will be
considered.

\section{Experimental details} \label{exp}

Single crystals of YTiO$_3$ were prepared by the floating zone
method using a four-mirror-type infrared image furnace from
Crystal System Corporation. More details about the crystal growth
are given in Ref. \onlinecite{kovaleva07}. Two samples have been
investigated and the measurements presented here were obtained on
the sample with the sharpest transition at $T_C$. This sample was
cut so that its faces are perpendicular to the $a$-, $b$-, and
$c$-axes, its dimensions at room temperature being equal to
$L_a^0\simeq2$ mm, $L_b^0\simeq4$ mm, and $L_c^0\simeq3$ mm along
$a$, $b$, and $c$, respectively, with a mass of 116 mg. Thermal
expansion and magnetostriction were measured using a home-made
high-resolution capacitive dilatometer \cite{meingast90,pott83},
with temperature and field sweep rates of 20 mK/s and 0.5 T/min,
respectively. Three sets of measurements were performed, where the
length $L_i$ was measured along the $a$-, $b$-, and $c$-axes
($i=a$, $b$, and $c$, respectively). Specific heat and
magnetization were measured using a Physical Properties
Measurement System and a Magnetic Properties Measurement System,
respectively (Quantum Design). For all measurements, the magnetic
field $\mathbf{H}$ was applied parallel to the easy axis
$\mathbf{c}$. The thermal expansion of EuS was measured using a 8
mm long single crystal grown from the melt by K. Fischer at the
Forschungszentrum J\"{u}lich, as described elsewhere
\cite{kobler75}.

\section{Results}
\label{thermodyn}

\subsection{Specific heat and thermal expansion}
\label{specthexp}

\begin{figure}[b]
    \centering
    \epsfig{file=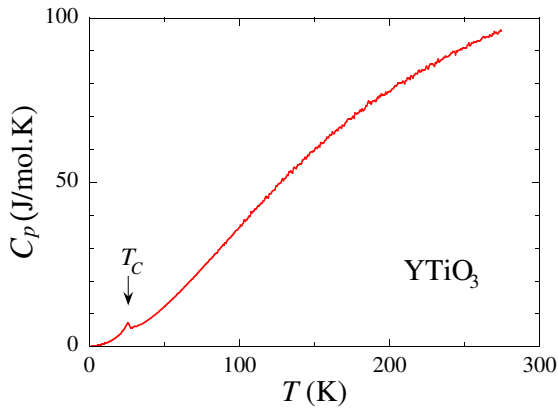,width=73mm}
    \caption{Variation with $T$ of the specific heat $C_p$ of YTiO$_3$.}
    \label{speheat}
\end{figure}

\begin{figure}[t]
    \centering
    \epsfig{file=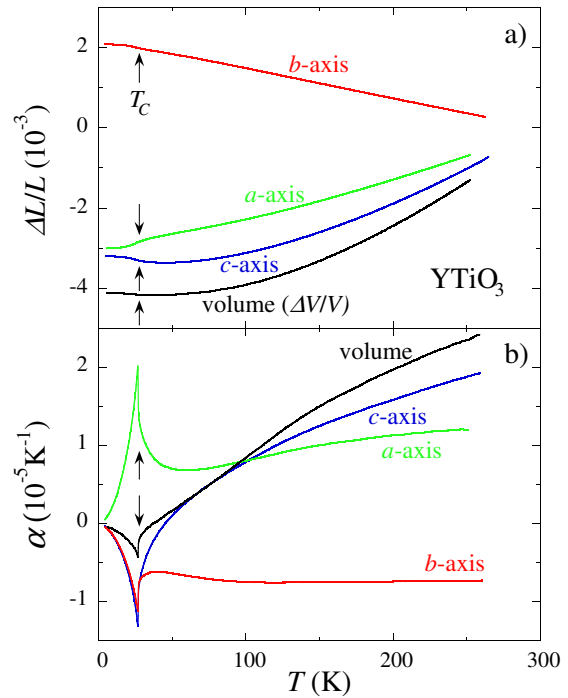,width=73mm}
    \caption{Variation with $T$ (a) of the relative lengths
    $\Delta L_i/L_i$ for $i=a$, $b$, and $c$, and of
    the relative volume $\Delta V/V$, (b) of
    the thermal expansion coefficients $\alpha_i$ for $i=a$,
    $b$, $c$ and $V$.}
    \label{thermal_expansion}
\end{figure}

In Fig. \ref{speheat}, the specific heat $C_p$ of YTiO$_3$ is
shown in a $C_p$ versus $T$ plot. Ferromagnetic ordering is
characterized by an anomaly at $T_C=26.8\pm 0.3$ K, defined at the
minimum of slope of $C_p(T)$. In Fig. \ref{thermal_expansion} (a),
the variation with $T$ of the relative lengths $\Delta L_i/L_i$ is
shown for $i=a$, $b$, and $c$, $\Delta L_i/L_i$ being fixed to
zero at room temperature. The linear thermal expansion
coefficients $\alpha_i=(1/L_i)\partial L_i/\partial T$, with
$i=a$, $b$, and $c$, are extracted from these data and are plotted
in Fig. \ref{thermal_expansion} (b). The volume change $\Delta
V/V=\sum_{i=a,b,c}\Delta L_i/L_i$ and the related volume thermal
expansion coefficient $\alpha_V=(1/V)\partial V/\partial T$ are
also shown in Fig. \ref{thermal_expansion} (a) and (b),
respectively. As seen in Fig. \ref{thermal_expansion}, changes of
slope in $L_a$, $L_b$, $L_c$, and $V$ are induced at $T_C$,
leading to a positive anomaly in $\alpha_a$ and to negative
anomalies in $\alpha_b$, $\alpha_c$, and $\alpha_V$. The volume
decrease below $T_C$, which is similar to the invar effect, will
be further related to the negative hydrostatic pressure dependence
of $T_C$. From the thermal expansion data, we extract a Curie
temperature $T_C=26.8\pm0.05$ K at the extremum of slope of
$\alpha_i(T)$. This is in good agreement with prior observations
\cite{kovaleva07}. At temperatures sufficiently higher than $T_C$,
$L_a$, $L_c$, and $V$ increase with $T$, while $L_b$ decreases
with $T$ (Fig. \ref{thermal_expansion} (a)). This leads to the
positive values of $\alpha_a$, $\alpha_c$, and $\alpha_V$ and to
the negative values of $\alpha_b$ observed at high temperatures in
Fig. \ref{thermal_expansion} (b). The anisotropy of $\alpha_i$ at
high temperatures is a consequence of the lattice distortions,
which will be discussed in Section \ref{discussion}.

\subsection{Magnetization and magnetostriction}
\label{magn}

\begin{figure}[t]
    \centering
    \epsfig{file=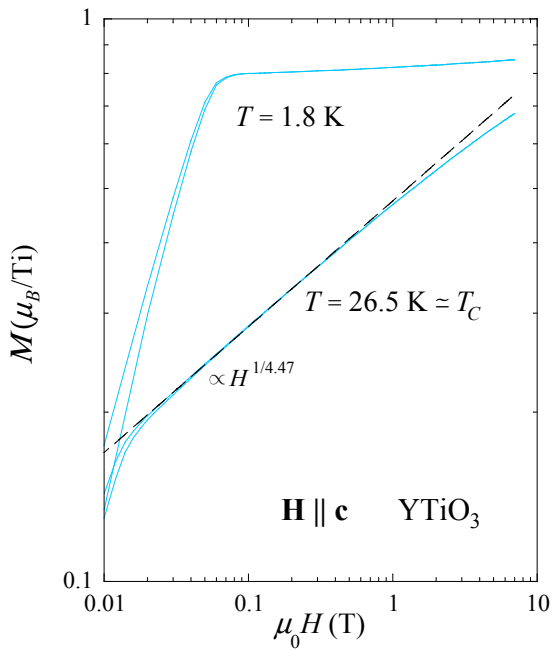,width=72mm}
    \caption{Field-dependence of the magnetization $M$, on a log-log
scale, at $T=$ 1.8 K and $T=26.5$ K $\simeq T_C$.}
    \label{magnetization}
\end{figure}

In Fig. \ref{magnetization}, the magnetization versus field $M(H)$
is shown for $\mathbf{H}\parallel\mathbf{c}$ in a log-log plot, at
$T=$ 1.8 K and $T=26.5$ K $\simeq T_C$. At $T=$ 1.8 K, a linear
increase of $M(H)$ is obtained for $\mu_0H<\mu_0H^*\simeq0.06$ T
and is related to the alignment of ferromagnetic domains. For
$H>H^*$, the spins are aligned parallel to $\mathbf{H}$ and the
magnetization $M$ reaches $M_s\simeq0.8$ $\mu_B$. In this regime,
a slight increase of $M(H)$ is observed. Indeed, $M_s$ is not yet
fully saturated and is somewhat smaller than the full moment of 1
$\mu_B$ expected for the $S=1/2$ Ti$^{3+}$ ions
\cite{ulrich02,tsubota00}. At $T\simeq T_C$, $M$ increases first
almost linearly with $H$, for $\mu_0H<\mu_0H^{*}{'}\simeq0.02$ T,
and then varies as $M\propto H^{1/\delta}$, with
$\delta=4.47\pm0.2$, for $\mu_0H^{*}{'}<\mu_0H<1$ T. This power
law, observed above 1 T, is characteristic of the critical
ferromagnetic regime. Deviations are observed when $M$ becomes
close to $M_s$.

In Ref. \onlinecite{kovaleva07}, low temperature magnetization
measurements were reported on the same sample as in the present
work and moments of 0.84 and 0.82 $\mu_B$ were found for a
magnetic field of 7 T applied along $\mathbf{b}$ and $\mathbf{c}$,
respectively. These data agree well with the saturated magnetic
moment of 0.84 $\mu_B$ reported in Ref. \onlinecite{tsubota00} and
\onlinecite{garret81}, and with the moments of 0.83 $\pm$ 0.05 and
0.84 $\pm$ 0.05 $\mu_B$, for the a- and c axis, respectively,
determined by recent Magnetic Compton Profile experiments
\cite{tsuji08}.

\begin{figure}[t]
    \centering
    \epsfig{file=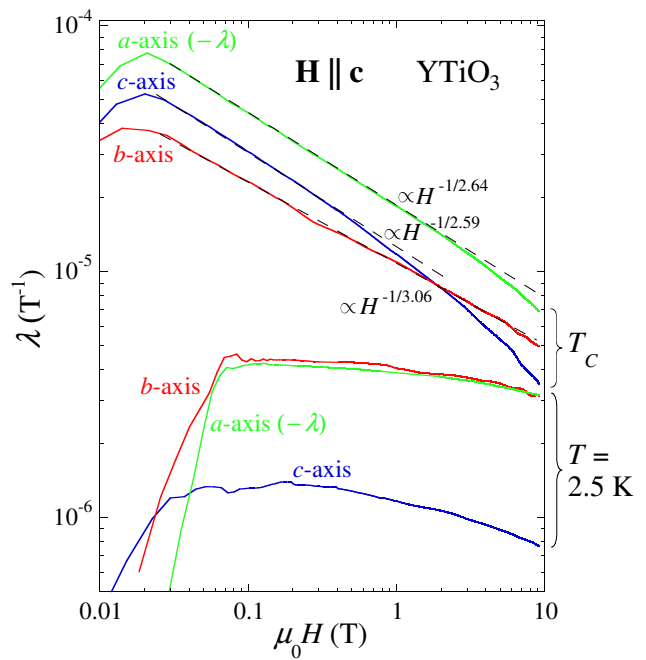,width=83mm}
    \caption{Field-dependence of the magnetostriction
    $\lambda_i$, on a log-log scale, for $i=a$,
    $b$, and $c$, at $T=2.5$ K and $T_C=26.7$ K
    ($\lambda_a$ is plotted with a minus sign).}
    \label{mgstr}
\end{figure}

In Fig. \ref{mgstr}, the magnetostriction coefficients
$\lambda_i=(1/L_i)\partial L_i/\partial (\mu_0H)$ are plotted as a
function of $H$ on a log-log scale, at $T=2.5$ K and $T_C=26.7$ K,
with $i=a$, $b$, and $c$, and $\mathbf{H}\parallel\mathbf{c}$. For
all temperatures and magnetic fields, $\lambda_b$ and $\lambda_c$
are positive while $\lambda_a$ is negative. For the three
configurations at $T=2.5$ K, $\left|\lambda_i(H)\right|$ increases
for $\mu_0H<\mu_0H^*\simeq0.06$ T and is almost constant for
$H>H^*$, when the domains are aligned. At $T_{C}$,
$\left|\lambda_i(H)\right|$ increases before reaching a maximum at
$\mu_0H^{*}{'}\simeq0.02$ T. For $\mu_0H>\mu_0H^{*}{'}$, a
critical regime is observed, where
$\left|\lambda_i(H)\right|\propto H^{-1/\delta'_i}$, with
$\delta'_i=$ 2.64, 3.06, and $2.59\pm0.1$ for $i=a$, $b$, and $c$,
respectively. While the power law is followed up to almost 10 T in
$\lambda_b$, deviations are found for $\mu_0H>1$ T in $\lambda_a$
and $\lambda_c$.

\section{Ferromagnetic properties} \label{analysis}

\subsection{Low temperature spin waves} \label{spinwaves}

\begin{figure}[t]
    \centering
    \epsfig{file=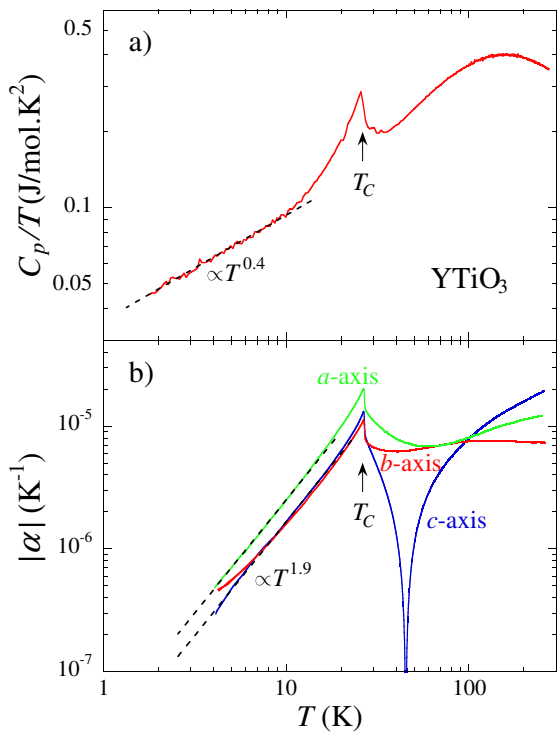,width=73mm}
    \caption{(a) Specific heat and (b) thermal expansion of
    YTiO$_3$, in $C_p/T$ and $|\alpha_i|$ versus $T$ on log-log scales, respectively.}
    \label{loglog}
\end{figure}

In Fig. \ref{loglog} (a), the specific heat of YTiO$_3$ is plotted
in a log-log plot of $C_p/T$ versus $T$. Well below $T_C$, the
phonon contribution can be neglected and the signal, which varies
as $C_p(T)\propto T^{1.4}$ up to 10 K, is believed to be only
magnetic. This power law is compatible with isotropic 3D
Heisenberg ferromagnetic spin waves, for which a $T^{1.5}$ law
would be expected, and is thus in good agreement with the
spin-wave dispersion observed by neutron scattering
\cite{ulrich02}. The slight deviation from the $T^{1.5}$ law might
result from a small spatial anisotropy of the exchange, the
spin-wave contribution to the specific heat of a ferromagnet
varying as $C_p(T)\propto T^{d/2}$, where $d$ is the
dimensionality of the exchange. It may also be related to possible
additional antiferromagnetic spin fluctuations and/or to spin
anisotropies originating from spin-orbit coupling (see below).

In Fig. \ref{loglog} (b), the thermal expansion is plotted in a
log-log plot of $|\alpha_i|$ versus $T$, for $i=a$, $b$, and $c$.
Power laws $\alpha_i\propto T^{1.9}$ are found up to almost 20 K
for $i=a$ and $c$, while no clear power law is observed for $i=b$.
As simple 3D Heisenberg ferromagnetic spin waves should lead to
$\alpha_i\propto C_p \propto T^{1.5}$, the different
$T$-dependences of $C_p$ and $\alpha_i$ reported here may result
from anisotropic exchange interactions, which lead to weak
additional magnetic Bragg reflections due to canting of the
ferromagnetic moments \cite{ulrich02}. Weak low-energy spin
fluctuations around these wave vectors may contribute to the
deviation of the temperature dependence of the low-temperature
specific heat and thermal expansion from the predictions of a
simple ferromagnetic Heisenberg model \cite{notespinwave}. Further
studies of the spin wave spectra (e.g. by neutron scattering) and
detailed calculations are needed for a quantitative explanation of
the results obtained here. Systematic studies by specific heat and
thermal expansion of the ATiO$_3$ family, such as the work
initiated in Ref. \onlinecite{komarek07}, may be of importance to
understand the evolution of the low temperature magnetic
properties.

\subsection{Arrott plot and critical fluctuations} \label{criticalfluctuations}

\begin{figure}[b]
    \centering
    \epsfig{file=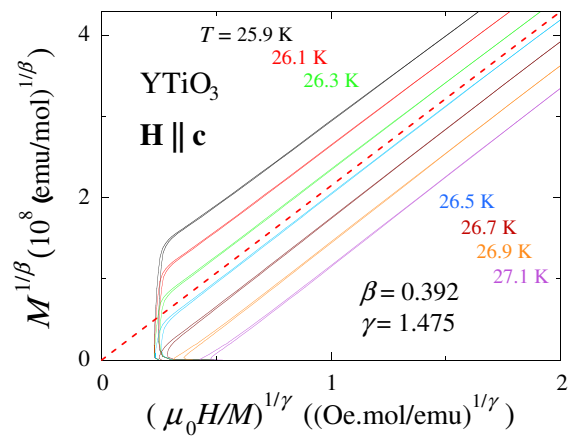,width=74mm}
    \caption{Best Arrott plot of the data,
    with $\beta=0.392$, $\gamma=1.475$, and $T_C=26.44$ K. The dotted
    line indicates the critical regime at $T_C$, associated with the exponent
    $\delta=4.76$.}
    \label{arrottplot}
\end{figure}

To analyze the critical ferromagnetic regime, the magnetization
was measured as $M$ versus $H$ at several temperatures close to
$T_C$. In Fig. \ref{arrottplot}, an Arrott plot of these data is
shown as $M^{1/\beta}$ versus $(H/M)^{1/\gamma}$, for $25.9\leq
T\leq27.1$ K. The critical exponents $\beta=0.392\pm0.05$ and
$\gamma=1.475\pm0.1$ used in this plot were determined from a fit
of $M(H,T)$ using Arrott's equation of state
\cite{arrott67,notecriticalexponents}:
\begin{eqnarray}
M^{1/\beta}=c_1\left(\frac{H}{M}\right)^{1/\gamma}-c_2\left(T-T_C\right),
    \label{arrottequation}
\end{eqnarray}
for $0.02<\mu_0H<1.3$ T and $25.9\leq T\leq27.1$ K. The value of
$T_C=26.44\pm0.05$ K obtained from this fit agrees well with that
obtained from the specific heat and thermal expansion (Section
\ref{specthexp}). From Equation (\ref{arrottequation}):
\begin{eqnarray}
&M(H,T_C)\propto H^{1/\delta},&\\
\rm{with}\;\;\;\;\;\;\;\;\;\;\;\;\;\;\;\;\;\;&\delta=(\beta+\gamma)/\beta.&\;\;\;\;\;\;\;\;\;\;\;\;\;\;\;\;\;\;\;\;\;\;\;\;\;\;\;\;\;\;\;
    \label{magnTC}
\end{eqnarray}
Using the exponents $\alpha$ and $\beta$ obtained with Arrott's
method, we calculate the critical exponent $\delta=4.76\pm0.2$,
which agrees favorably with $\delta=4.47\pm0.2$ directly obtained
from the fit by a power law of $M(H)$ at 26.5 K (Section
\ref{magn}).

In Table \ref{table1}, the exponents $\beta$, $\gamma$, and
$\delta$ expected for different classes of universality
\cite{collins89} are listed for comparison. The exponents
extracted from the Arrott plot of the magnetization of YTiO$_3$
are rather close to those of the 3D Heisenberg universality class.
A similar plot was made in Ref. \onlinecite{cheng08} using 3D
Heisenberg exponents, but without a preliminary fit of the
$M(H,T)$ data as done here. As already inferred from the behavior
of the low temperature spin waves, the critical behavior of the
magnetization, too, is thus consistent with a 3D Heisenberg
picture of ferromagnetism (cf. Section \ref{spinwaves} and Ref.
\onlinecite{ulrich02}).

In the following, the power laws reported for the magnetostriction
at $T_C$ (Fig. \ref{mgstr}) are related to the critical power law
of the magnetization (Fig. \ref{magnetization}) and to the
critical exponents $\alpha$ and $\beta$. Using the Maxwell
relation:
\begin{eqnarray}
\lambda_i=\frac{1}{L_i}\frac{\partial L_i}{\partial
(\mu_0H)}=-\frac{\partial M}{\partial p_i},
    \label{maxwell}
\end{eqnarray}
the magnetostriction coefficients can be expressed as functions of
the uniaxial pressure dependences of the magnetization. Assuming
that $\partial c_1/\partial p_i=0$, the derivative of Equation
(\ref{arrottequation}) leads to, at $T=T_C$:
\begin{eqnarray}
&\;\;\;\lambda_i(H,T_C)=-{\displaystyle A\frac{\partial
T_C}{\partial p_i}H^{-1/\delta'}},&\\
\rm{with}&\;\;A=c_1^{-\gamma/\delta'}c_2 \gamma/\delta \;\;\;
\rm{and}\;\;\;\delta'=(\beta+\gamma)/(1-\beta).&\;\;\;\;\;
   \label{Dershort}
\end{eqnarray}
From the exponents $\alpha$ and $\beta$ obtained by the Arrott fit
of the magnetization, a critical exponent $\delta'= 3.07\pm0.5$ is
expected to characterize the magnetostriction at $T_C$. This value
is in good agreement with the values $2.59\leq\delta'_i\leq3.06$
determined from the fits of $\lambda_i(H)$ at $T_C$ (see Section
\ref{magn}). For each set of ($\beta$,$\gamma$), the corresponding
$\delta'$ values are also given in Table \ref{table1}. These
values are compatible with a 3D Heisenberg scenario of
ferromagnetism for YTiO$_3$ associated with $\delta'= 2.77$. The
slight variations of $\delta_i'$ with $i$ are not understood and
may result from various secondary effects (anisotropic energy
scales, defects etc.).

To our knowledge, YTiO$_3$ is the second ferromagnetic system
known, after the itinerant ferromagnet UIr \cite{knafo08a}, where
a critical power law is reported in the magnetostriction at $T_C$.
We believe that such an effect is quite general and should be
present in most ferromagnets, once the field and temperature
ranges are properly chosen. We note that, more than 60 years ago,
Belov has theoretically predicted a similar law \cite{belov56},
but only within a mean-field approach, which corresponds to
$\beta=0.5$, $\gamma=1$, and $\delta=\delta'=3$ (cf. Table
\ref{table1}). Our approach is more general and permits to obtain
the critical exponent $\delta'$ for each combination of
($\beta$,$\gamma$) and thus, for each universality class.

\begin{table}[t] \caption{Critical exponents
$\beta$, $\gamma$, $\delta$, and $\delta'$ obtained here for
YTiO$_3$ and expected for different universality classes
\cite{collins89}.}
\begin{ruledtabular}
\begin{tabular}{lcccc}
Critical exponents&$\beta$&$\gamma$&$\delta$&$\delta'$\\
\hline
YTiO$_3$ (best fit) &0.392 (50)&1.475 (100)&4.76 (20)&3.07 (50)\\
3D Heisenberg&0.367&1.388&4.78&2.77\\
3D XY&0.345&1.316&4.81&2.54\\
3D Ising&0.326&1.238&4.80&2.32\\
2D Ising&0.125&1.75&15&2.14\\
Mean Field&0.5&1&3&3
\end{tabular}
\end{ruledtabular}
\label{table1}
\end{table}

\subsection{High-temperature magnetic signal: deviation from a pure
3D Heisenberg ferromagnet} \label{hightemp}

\begin{figure}[b]
    \centering
    \epsfig{file=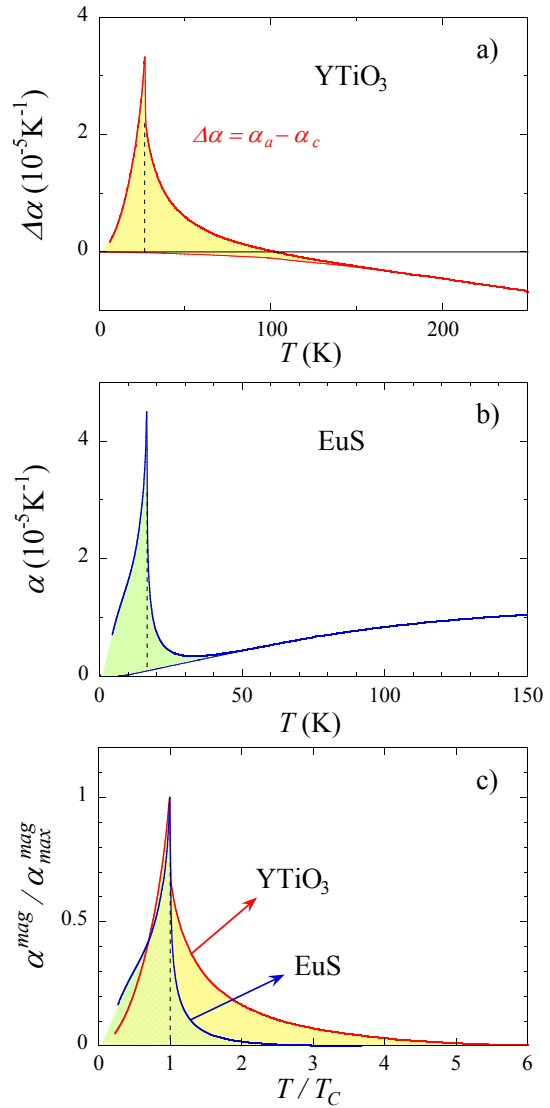,width=71mm}
    \caption{Temperature dependence of (a) $(\Delta\alpha=\alpha_a-\alpha_c)$ for
    YTiO$_3$, and of (b) $\alpha$ for EuS. In both plots, the thin line is a guide to the eye
    indicating the non-magnetic background, the colored area is
    the magnetic contribution deduced from the raw data after substraction of this background, and the
    vertical dotted line indicates the ferromagnetic transition temperature.
    (c) Magnetic contribution to the thermal expansion estimated for YTiO$_3$ (red line)
    and for EuS (blue line), in a normalized $\alpha^{mag}/\alpha^{mag}_{max}$ vs $T/T_C$ plot.}
    \label{thexpdiff}
\end{figure}

The distortion of the lattice induces the strong anisotropy
observed in the high-temperature thermal expansion of YTiO$_3$
(see Fig. \ref{thermal_expansion}). This anisotropic lattice
signal must be properly taken into account to extract the magnetic
contribution to the thermal expansion. Since $\alpha_a$ and
$\alpha_c$ are rather close above 100 K, we assume that the
lattice contributions to $\alpha_a$ and $\alpha_c$ are similar, so
that $\Delta\alpha=\alpha_a-\alpha_c$, which is plotted in Fig.
\ref{thexpdiff} (a), can be considered as a signal representative
of the magnetic thermal expansion. In this plot, the thin red line
is a guide to the eye indicating the "remaining" non-magnetic
background and the yellow area corresponds to the estimated
magnetic contribution. In Fig. \ref{thexpdiff} (b), the thermal
expansion $\alpha$ of cubic EuS \cite{notecubicalpha}, which is
known as a prototype of 3D Heisenberg ferromagnetism
\cite{kornblit78,wosnitza89,boni95}, is shown for comparison. The
thin blue line is a guide to the eye indicating the non-magnetic
background and the magnetic contribution is estimated by the blue
area. The estimates of the magnetic contribution to the thermal
expansion of YTiO$_3$ and EuS are plotted in Fig. \ref{thexpdiff}
(c). In this plot, the magnetic thermal expansion coefficient
$\alpha^{mag}$ is normalized by its maximal value
$\alpha^{mag}_{max}$ and the temperature $T$ is normalized by
$T_C$. Fig. \ref{thexpdiff} (c) indicates that, in YTiO$_3$, the
magnetic signal has a significant weight above $T_C$ and extends
up to about $5\times T_C$ while, in EuS, it has most of its
intensity below $T_C$ and vanishes completely above about $2\times
T_C$. Thus, the magnetic fluctuations of YTiO$_3$ cannot be
described as those of a simple 3D Heisenberg ferromagnet.

\begin{figure}[t]
    \centering
    \epsfig{file=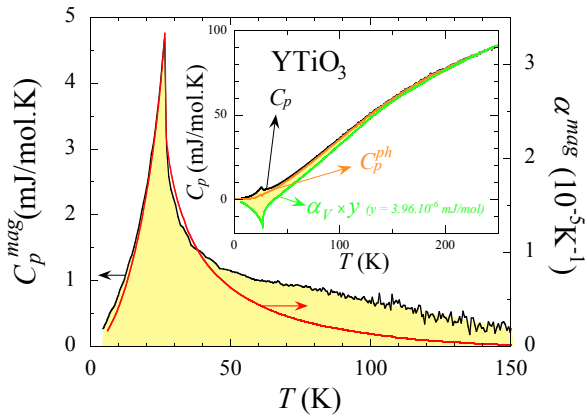,width=77mm}
    \caption{Temperature dependence of the magnetic contribution $C_p^{mag}$ to the specific
    heat, in a $C_p^{mag}/T$ versus $T$ plot. The inset shows the
    scaling of $C_p(T)$ (red line) and $\alpha_V(T)$ (green line), which permitted to
    estimate the non-magnetic contribution (blue line).}
    \label{specheatmag}
\end{figure}

As shown in the inset of Fig. \ref{specheatmag}, the specific heat
$C_p(T)$ and the volume thermal expansion $\alpha_V(T)$ can be
scaled at high temperatures using an empirical parameter
$y=3.96\times10^{-6}$ mJ/mol defined by $C_p=y\times\alpha_V$.
Assuming that the magnetic contribution to $C_p$ and $\alpha_V$ is
negligible above 200 K, and that there is a single Gruneisen
parameter associated with the phonons for the [0,300 K] range, we
can estimate the phonon contribution to the specific heat by
$C_p^{ph}=C_p-z\times(C_p-y\times\alpha_V)$, $z=0.2$ being
adjusted so that no anomaly remains at $T_C$. The main frame of
Fig. \ref{specheatmag} shows the resulting estimate of the
magnetic contribution $C_p^{mag}(T)$ to the specific heat. This
plot confirms the conclusions from Fig. \ref{thexpdiff}, i.e. that
a magnetic contribution is present up to more than 100 K.
Integration of the estimated magnetic heat capacity leads to the
magnetic entropy $\Delta S^{mag}\simeq4.5$ J/mol.K. This entropy
is roughly equal to the full spin entropy $\Delta
S^{mag}_{full}=R\rm{ln}2\simeq5.8$ J/mol.K expected for $S=1/2$
spin system; the fact that $\Delta S^{mag}$ is about 20 \% smaller
than $\Delta S^{mag}_{full}$ may be imputed to the experimental
error.

In Fig. \ref{specheatmag}, the estimated magnetic contributions to
the thermal expansion $\alpha^{mag}(T)$ and to the specific heat
$C_p^{mag}(T)$ are plotted together; above 50 K, the different
shapes of the signals indicate the limit of the methods used here.
Both plots indicate clearly the presence of a magnetic signal at
temperatures well above ferromagnetic ordering. The origin of this
behavior, which is not expected for usual 3D-Heisenberg
ferromagnets (see Fig. \ref{thexpdiff}), is not yet understood. In
principle, a modified conventional spin-only fluctuation model,
for example with competing (and possibly low-dimensional)
antiferromagnetic and ferromagnetic interactions, could describe
this anomalously high-temperature magnetic signal. However, this
is in apparent contradiction with the magnon spectra reported by
neutron scattering, which do not exhibit pronounced deviations
from the predictions of a 3D Heisenberg model with
nearest-neighbor interactions \cite{ulrich02}. An alternative
explanation of the extended magnetic fluctuation regime could be
offered by spin-orbital fluctuations models, where an energy scale
significantly exceeding the magnon bandwidth (in Ref.
\onlinecite{ulrich06,ulrich08}, orbital fluctuations were
associated with an excitation at about 250 meV) could actuate
ferromagnetic fluctuations at temperatures between $T_C$ and room
temperature \cite{khaliullin03}. Further work is required to
ascertain whether a quantitatively consistent picture of the spin
\cite{ulrich02} and orbital \cite{ulrich06,ulrich08} excitation
spectra and thermodynamics of YTiO$_3$ can be obtained.

\section{Coupling between the lattice and the magnetic
properties}\label{discussion}

\subsection{Uniaxial pressure dependences - Comparison with LaTiO$_3$}
\label{uniaxialpress}

\begin{table}[b]
\caption{Uniaxial and hydrostatic pressure dependences of $T_C$
and $M_s$ for YTiO$_3$ and of $T_N$ for LaTiO$_3$. The ratio
$\rho_i$ of the pressures dependences of $T_C$ and $M_s$ is given
for YTiO$_3$.}
\begin{ruledtabular}
\begin{tabular}{ccccc}
\multicolumn{1}{c}{} & \multicolumn{3}{c}{YTiO$_{3}$} &  \multicolumn{1}{c}{LaTiO$_{3}$} \\
&$\partial \rm{ln}T_{C}/\partial p_i$&$\partial  \rm{ln}M_s/\partial p_i$&$\rho_i$&$\partial  \rm{ln}T_{N}/\partial p_i$\\
& (10$^{-3}$kbar$^{-1}$)& (10$^{-3}$kbar$^{-1}$)&& (10$^{-2}$kbar$^{-1}$)\\
\hline
$p_{a}$& 9.9 $\pm$ 1.0&2.7 $\pm$ 0.4&3.7 $\pm$ 1.0&-1.9 $\pm$ 0.4\\
$p_{b}$& -5.1 $\pm$ 0.5&-2.7 $\pm$ 0.4&1.9 $\pm$ 0.5&1.9 $\pm$ 0.4\\
$p_{c}$& -7.1 $\pm$ 0.7&-0.77 $\pm$ 0.15&9.1 $\pm$ 2.5&$\simeq0$\\
$p_{h}$& -2.3 $\pm$ 0.3&-0.77 $\pm$ 0.15&2.9 $\pm$ 1.0&$\simeq0$
\end{tabular}
\end{ruledtabular}
\label{table2}
\end{table}

\begin{figure}[t]
    \centering
    \epsfig{file=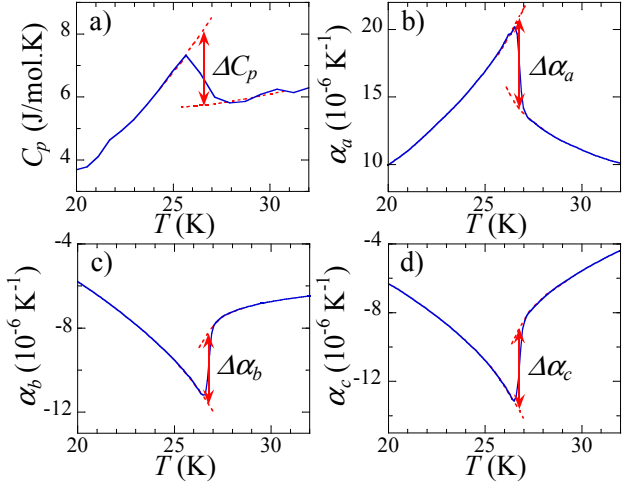,width=82mm}
    \caption{Ferromagnetic ordering anomaly (a) in the specific heat
and (b-d) in the thermal expansivity coefficients along $a$, $b$,
and $c$. In these curves, the jumps $\Delta C_p$,
$\Delta\alpha_a$, $\Delta\alpha_b$, and $\Delta\alpha_c$ at the
ferromagnetic ordering are indicated by arrows.}
    \label{ehrenfest}
\end{figure}

Fig. \ref{ehrenfest} shows the anomalies at $T_C$ in the specific
heat and in the thermal expansion of YTiO$_3$. These anomalies are
typical of a second-order phase transition, whose jumps are
estimated as $\Delta C_p=2.3\pm0.1$ J/molK in the specific heat
and as $\Delta\alpha_a=6.6\pm0.3\times10^{-6}$ K$^{-1}$,
$\Delta\alpha_b=-3.4\pm0.2\times10^{-6}$ K$^{-1}$, and
$\Delta\alpha_c=-4.7\pm0.2\times10^{-6}$ K$^{-1}$ in the thermal
expansion. Using the Ehrenfest relation:
\begin{eqnarray}
\frac{\partial T_C}{\partial
p_i}=\frac{\Delta\alpha_iV_mT_C}{\Delta C_p},
    \label{ehrenfestrelation}
\end{eqnarray}
where $V_m=3.46\times10^{-5}$ m$^3$/mol is the molar volume and
$p_i$ a uniaxial pressure applied along $i$ ($i=a$, $b$, and $c$),
we extract the uniaxial pressure dependences $\partial
T_C/\partial p_i$ reported in Table \ref{table2}. The sum of the
three uniaxial pressure dependences of $T_C$ gives the hydrostatic
pressure dependence $\partial T_C/\partial
p_h=-6.0\pm0.6\times10^{-2}$ K/kbar. Assuming that, in YTiO$_3$,
$\partial T_C/\partial p_h$ remains constant under hydrostatic
pressure, ferromagnetism may be destroyed above $p_c\simeq 400$
kbar.

For comparison, the uniaxial pressure dependences of $T_N$, for
the antiferromagnet LaTiO$_3$, are also listed in Table
\ref{table2}. To calculate them, $T_C$ in Equation
(\ref{ehrenfestrelation}) was substituted by $T_N\simeq146$ K,
$\Delta C_p=10$ J/molK, $\Delta\alpha_a=-5\pm0.5\times10^{-5}$
K$^{-1}$, $\Delta\alpha_b=5\pm0.5\times10^{-5}$ K$^{-1}$, and
$\Delta\alpha_c\simeq0$ being estimated from Ref.
\onlinecite{hemberger03}. While $\partial T_{N}/\partial p_c$ is
too small to be extracted from Ref. \onlinecite{hemberger03},
$\partial T_{N}/\partial p_a$ and $\partial T_{N}/\partial p_b$
are such that $\partial T_{N}/\partial p_a=-\partial
T_{N}/\partial p_b<0$. For
$\mathbf{p}\parallel\mathbf{a},\mathbf{b}$, the uniaxial pressure
dependences of $T_C$ and $T_N$, for YTiO$_3$ and LaTiO$_3$,
respectively, have thus opposite sign. The effects of pressure
along $c$ are such that $\partial T_{C}/\partial p_c<0$ and
$\partial T_{N}/\partial p_c\simeq0$. In Sections \ref{GdFeO3} and
\ref{jahnteller}, the uniaxial pressure-dependences of $T_C$ and
$T_N$ in YTiO$_3$ and LaTiO$_3$, respectively, will be interpreted
as resulting from pressure-induced modifications of the
distortion.

Well below $T_C$ and for $\mu_0H>\mu_0H^*=0.06$ T, $M\simeq
M_s\simeq$ 0.8 $\mu_B$ (Fig. \ref{magnetization}) and the
magnetostriction coefficients of YTiO$_3$ are almost constant,
having the values $\lambda_a\simeq-3.5\pm0.5\times10^{-6}$
T$^{-1}$, $\lambda_b\simeq3.5\pm0.5\times10^{-6}$ T$^{-1}$, and
$\lambda_c\simeq1.0\pm0.2\times10^{-6}$ T$^{-1}$ (Fig.
\ref{mgstr}). Using the Maxwell relation given in Eq.
(\ref{maxwell}), we extract:
\begin{eqnarray}
\frac{\partial M_s}{\partial p_i}=-\lambda_i(T=2.5\;\rm{K}).
    \label{dMsdpmaxwell}
\end{eqnarray}
The values of $\partial M_s/\partial p_i$, for $i=a$, $b$, and
$c$, as well as their sum, the hydrostatic pressure dependence
$\partial M_s/\partial p_h$, are summarized in Table \ref{table2}.

For each $i=a$, $b$, $c$, and $h$ ($h\leftrightarrow$
hydrostatic), the pressure dependences $\partial T_{C}/\partial
p_i$ and $\partial M_s/\partial p_i$ have always the same sign,
being both positive for $i=a$ and both negative for $i=b,c,h$
(Table \ref{table2}). Ferromagnetic order is thus stabilized by
uniaxial pressure $\mathbf{p}\parallel\mathbf{a}$ and is
destabilized by uniaxial pressure
$\mathbf{p}\parallel\mathbf{b},\mathbf{c}$ and by hydrostatic
pressure. Consequently, the ratio $\rho_i$, defined by:
\begin{eqnarray}
\rho_i=\frac{M_s}{T_C}\frac{\partial T_C/\partial p_i}{\partial
M_s/\partial p_i},
    \label{ratio}
\end{eqnarray}
is always positive. As shown in Table \ref{table2}, we find that
$\rho_i$ is strongly anisotropic, being bigger when the $i$-axis
is easier ($c=$ easy, $a=$ intermediate, and $b=$ hard
\cite{tsubota00,kovaleva07}).

Although YTiO$_3$ is a localized ferromagnet, its saturated
moment, at about 5 T, is only 80 \% of the fully saturated moment
$M_{s}^{full}=1$ $\mu_B$. A small canted antiferromagnetic moment
$M_{AF}\approx0.1$ $\mu_B$ was reported by neutron scattering in
YTiO$_3$ \cite{ulrich02} and explains partly why $M_s$ is reduced.
In addition, the reduction of $M_s$ may indicate an enhanced phase
space for quantum magnetic fluctuations. The question is whether
this reduction comes from usual spin-only fluctuations, or if it
results from more complicated fluctuations involving orbital
degrees of freedom \cite{khaliullin03,ulrich02}. The high values
of $\partial M_s/\partial p_i$ in YTiO$_3$ may result from the
combination of two effects, which can be summarized as the
uniaxial pressure-induced transfers of weight (i) between the
ferromagnetic moment $M_s$ and the antiferromagnetic moment
$M_{AF}$ and, (ii) between $M_s$ and some quantum magnetic
fluctuations $\delta M$. The second effect is similar to what
happens in itinerant ferromagnets, where $M_s$ is reduced by
quantum fluctuations $\delta M$ of the magnetic moment, and where
the strong pressure dependences of $M_s$ are related to those of
$\delta M$.

A similar analysis as the one presented here was reported for the
itinerant weak ferromagnet UIr (Ref. \onlinecite{knafo08a}), in
the framework of the Moriya's spin fluctuation theory of itinerant
magnetism \cite{moriya95,takahashi06}. By analogy, a spin
fluctuation theory, adapted to the particular case of YTiO$_3$,
may be appropriate.

\subsection{Coupling between the GdFeO$_3$-type distortion and the magnetic ordering}
\label{GdFeO3}

\begin{figure}[t]
    \centering
    \epsfig{file=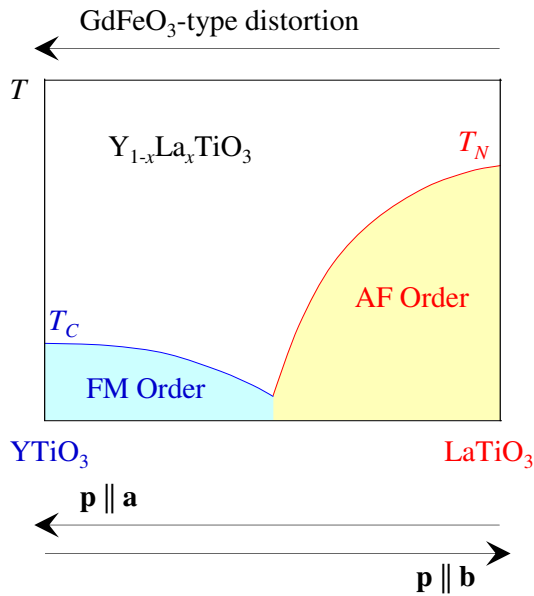,width=70mm}
    \caption{Schematic magnetic phase diagram of
    Y$_{1-x}$La$_x$TiO$_3$ (AF = antiferromagnetic and FM =
    ferromagnetic). The arrows indicate the effects of
    increasing the GdFeO$_3$-type
    distortion and of applying uniaxial pressures
    $\mathbf{p}\parallel\mathbf{a},\mathbf{b}$.}
    \label{phase_diagram}
\end{figure}

In the ATiO$_3$ perovskites, the GdFeO$_3$-type distortion
comprises a combination of tilts and rotations of the TiO$_6$
octaedra. This results in an orthorhombic structure, where
$b>a_0\sqrt{2}>c/\sqrt{2}>a$, $a_0$ being the lattice parameter of
an undistorted cubic structure \cite{duboulay95,notelatticecubic}.
In the alloys Y$_{1-x}$La$_x$TiO$_3$, La-substitution induces a
decrease of the GdFeO$_3$-type distortion, which is believed to
control the change from ferromagnetism to antiferromagnetism
\cite{goral82,okimoto95,goodenough04,mochizuki04,pavarini05} (Fig.
\ref{phase_diagram}). This picture, in which ferromagnetism is
favored by a large GdFeO$_3$-type distortion, is qualitatively
confirmed by the increase of the distortion of the ($a$,$b$) plane
induced below $T_C$ (see Fig. \ref{thermal_expansion}). In the
following, we will further show that, for
$\mathbf{p}\parallel\mathbf{a},\mathbf{b}$, the uniaxial pressure
dependences of $T_C$ and $T_N$ are mainly controlled by those of
the GdFeO$_3$-type distortion.

In YTiO$_3$ and LaTiO$_3$, the negative sign of $\partial
T_{C}/\partial p_b$ and the positive sign of $\partial
T_{N}/\partial p_b$ (see Table \ref{table2}), respectively, imply
that a uniaxial pressure $\mathbf{p}\parallel\mathbf{b}$ can be
seen as equivalent to La-doping (cf. the corresponding arrow in
Fig. \ref{phase_diagram}). Conversely, the fact that $\partial
T_{C}/\partial p_a$ is positive while $\partial T_{N}/\partial
p_a$ is negative (Table \ref{table2}) implies that
$\mathbf{p}\parallel\mathbf{a}$ is equivalent to Y-doping (see
Fig. \ref{phase_diagram}). As $\mathbf{p}\parallel\mathbf{b}$
induces a compression along $b$ and, because of elasticity,
extensions along $a$ and $c$, its effects are very similar to
those of reducing the GdFeO$_3$-type distortion. By analogy,
$\mathbf{p}\parallel\mathbf{a}$ leads to a compression along $a$
and to small extensions along $b$ and $c$, which is similar to
increasing the GdFeO$_3$-type distortion. Thus, we conclude that
$\mathbf{p}\parallel\mathbf{a}$ and
$\mathbf{p}\parallel\mathbf{b}$ induce an increase and a decrease
of the GdFeO$_3$-type distortion, respectively, which are
responsible for the various signs of $\partial T_{C,N}/\partial
p_i$, for $i=a,b$.

The uniaxial pressures dependences of $T_C$ and $T_N$ are a
consequence of the high sensitivity of the superexchange
interactions to the bond angles between the ions, whose positions
are very sensitive to the pressure-induced modifications of the
GdFeO$_3$-type distortion. The application of uniaxial pressures
$\mathbf{p}\parallel\mathbf{a},\mathbf{b}$, as well as the
variation of the A$^{3+}$ ion, permits thus to tune the
competition between the ferromagnetic and the antiferromagnetic
exchange interactions, via a change of the GdFeO$_3$-type
distortion.

However, a pressure-driven change of the GdFeO$_3$-type distortion
cannot explain the results obtained for
$\mathbf{p}\parallel\mathbf{c}$, i.e. $\partial T_{C}/\partial
p_c<0$ and $\partial T_{N}/\partial p_c\simeq0$. Indeed, $\partial
T_{C}/\partial p_c>0$ and $\partial T_{N}/\partial p_c<0$ would be
expected if $\mathbf{p}\parallel \mathbf{c}$ merely modified the
GdFeO$_3$-type distortion (since $\mathbf{p}\parallel \mathbf{c}$
contracts $c$, it should increase the GdFeO$_3$-type distortion).
Another mechanism, in addition to the GdFeO$_3$-type distortion,
is needed to understand the pressure dependences of $T_C$ and
$T_N$ for $\mathbf{p}\parallel\mathbf{c}$. In the next Section, we
will show that a higher sensitivity of the $c$-axis length to the
intrinsic elongations of the octahedra may be the origin of this
behavior.

\subsection{Distortion of the TiO$_6$ octahedra}
\label{jahnteller}

\subsubsection{Microscopic description}
\label{microscopicdescription}

\begin{figure}[b]
    \centering
    \epsfig{file=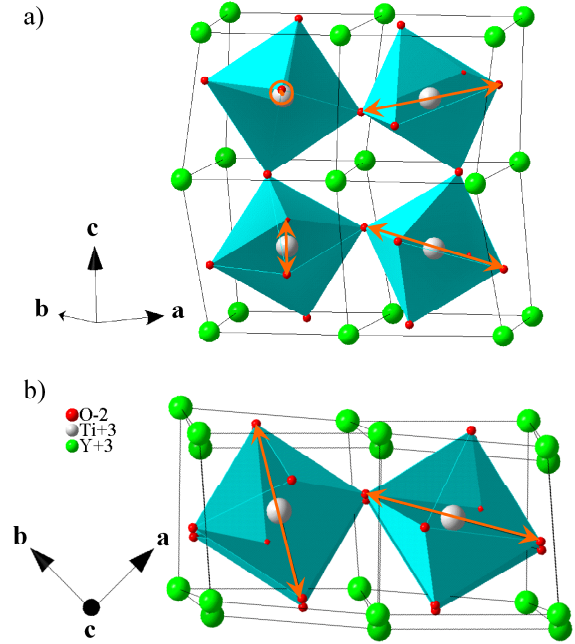,width=75mm}
    \caption{Schematics of the lattice structure of YTiO$_3$. The Ti$^{3+}$ ions
are represented by grey spheres, the Y$^{3+}$ ions by green
spheres, and the O$^{2-}$ ions by red spheres. The TiO$_6$
octahedra are colored in blue, and orange arrows show their
elongated direction, possibly due to the Jahn-Teller distortion.}
    \label{structure_JT_distortion}
\end{figure}

The lattice structure of YTiO$_3$ is represented schematically in
Fig. \ref{structure_JT_distortion}, where the alternation of tilts
and rotations of the TiO$_6$ octahedra (in blue) is due to the
GdFeO$_3$-type distortion. An additional distortion consists of an
elongation of each octahedron along a particular axis (orange
arrows in Fig. \ref{structure_JT_distortion}), and of contractions
perpendicularly to this axis. In Ref.
\onlinecite{mochizuki04,pavarini05,akimitsu01,iga04}, the fact
that the elongated axes vary from one site to another was ascribed
to a staggered ordering of the $t_{2g}$ orbitals (Ti$^{3+}$ ions)
via a collective Jahn-Teller effect. By geometrical
considerations, we can qualitatively estimate the macroscopic
distortion induced by the elongations of the octahedra. As seen in
Fig. \ref{structure_JT_distortion}, the elongated axes are almost
contained within the ($a,b$) plane, i.e. perpendicularly to the
$c$-axis. This implies that the elongations of the octahedra
induce a contraction of the $c$-axis. In the ($a,b$) plane, the
elongated axes of two adjacent octahedra subtend an angle of about
60 $^{\circ}$, so that the elongations and contractions of the
different octahedra almost cancel each other. Since the
projections of the elongated axes are larger along $b$ than along
$a$ (the elongated axes subtend an angle of about 30 $^{\circ}$
with $b$ and of about 60 $^{\circ}$ with $a$), we finally conclude
that the elongations of the octahedra are responsible for a small
elongation of $b$ and for a tiny compression of $a$, in addition
to the main effect, a compression along $c$.

\subsubsection{Lattice parameters: comparison of the families ATiO$_3$ and AFeO$_3$}
\label{latticeparam}

Here we propose a method, based on a comparison of the lattice
parameters $a$, $b$ and $c$ of the families ATiO$_3$ and AFeO$_3$,
to confirm the description made in Section
\ref{microscopicdescription} of the effects of the elongation of
the octahedra on $a$, $b$ and $c$. Assuming that these elongations
are related to a Jahn-Teller distortion \cite{note_jahn_teller},
the comparison of the lattice parameters of ATiO$_3$ and AFeO$_3$
can be justified by the fact that, contrary to Ti$^{3+}$,
Fe$^{3+}$ is not Jahn-Teller active so that AFeO$_3$ can be
considered as a non-Jahn-Teller reference for ATiO$_3$.

\begin{widetext}

\begin{figure}[h]
    \centering
    \epsfig{file=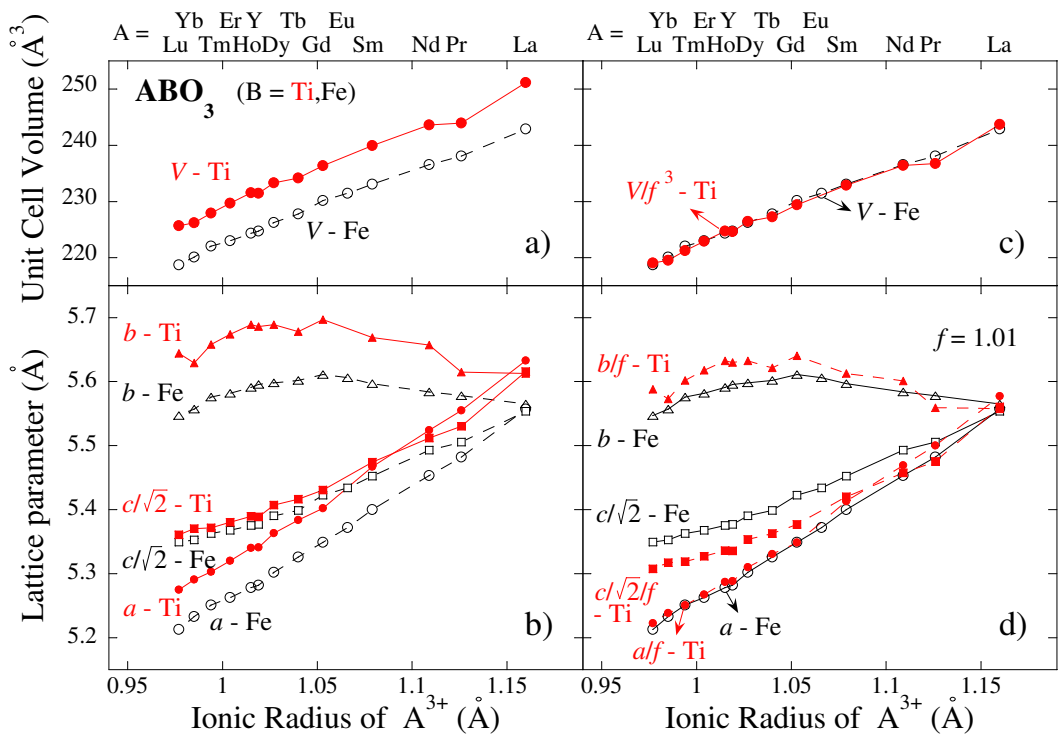,width=140mm}
    \caption{Variations, for the ATiO$_3$ and AFeO$_3$ systems, of
(a) the unit cell volume $V$ and (b) the lattice parameters $a$,
$b$, and $c/\sqrt{2}$, as a function of the ionic radius of
A$^{3+}$. In (c) and (d), the volume and the lattice parameters of
ATiO$_3$ and AFeO$_3$ are scaled together using the empiric factor
$f=1.01$.}
    \label{lenghtABO3}
\end{figure}
\end{widetext}

In Fig. \ref{lenghtABO3} (a) and (b), the unit cell volume $V$ and
the lattice parameters $a$, $b$, and $c/\sqrt{2}$ are plotted
versus the ionic radius of the A$^{3+}$ ions, for several
compounds of the families ATiO$_3$ and AFeO$_3$
(A=Lu$\rightarrow$La)
\cite{zhou05,maclean79,duboulay95,marezio70,shannon76}. The ionic
radii of the A$^{3+}$ ions are taken from Ref.
\onlinecite{shannon76}, assuming a number of 8 nearest neighbors
\cite{komarek07}. While LaTiO$_3$ and LaFeO$_3$ are almost
undistorted ($a\simeq b\simeq c/\sqrt{2}$), Fig. \ref{lenghtABO3}
(b) shows a strong distortion of the pseudo-cubic lattice in
ATiO$_3$ and AFeO$_3$, once A$^{3+}$ is smaller than La$^{3+}$.

In Fig. \ref{lenghtABO3} (c), the unit cell volumes
$V^{\rm{ATiO}_3}$ of the ATiO$_3$ compounds are scaled empirically
with the unit cell volumes $V^{\rm{AFeO}_3}$ of the AFeO$_3$
compounds, using a scaling factor $f=1.01$ defined by
$V^{\rm{ATiO}_3}=V^{\rm{AFeO}_3}*f^3$. In Fig. \ref{lenghtABO3}
(d), the lattice parameters of ATiO$_3$ are scaled to those of
AFeO$_3$ using the factor $1/f$. As the undistorted limit in the
ABO$_3$ perovskites corresponds to a cubic lattice parameter
$a_0^{\rm{ABO}_3}=2(r_{\rm{O}}+r_{\rm{B}})$, where $r_{\rm{O}}$
and $r_{\rm{B}}$ are the ionic radii of the O$^{2-}$ and B$^{3+}$
ions, respectively, we associate the empirical scaling factor
$f=1.01$ to the ratio $a_0^{\rm{ATiO}_3}/a_0^{\rm{AFeO}_3}=1.013$,
calculated with $r_{\rm{O}}=1.35$, $r_{\rm{Ti}}=0.67$, and
$r_{\rm{Fe}}=0.645$ \AA \cite{shannon76}.

Since the ionic radii of Ti$^{3+}$ and Fe$^{3+}$ are very close,
we assume that, for the two families, the GdFeO$_3$-type
distortion induces similar variations of their lattice parameters
in the scaled plot of Fig. \ref{lenghtABO3} (d). Consequently, the
elongations of the octahedra, which can be neglected in the
non-Jahn-Teller compound AFeO$_3$, might be responsible for the
slight differences, in Fig. \ref{lenghtABO3} (d), between the
scaled lattice parameters of the two families. This implies that
the elongations of the TiO$_6$ octahedra in YTiO$_3$ induce a
decrease of $c/\sqrt{2}$ by about 0.5 $\rm{\AA}$, accompanied by a
smaller increase of $b$, by about 0.3 $\rm{\AA}$, and by no
noticeable change of $a$. These conclusions, obtained using the
scaled plot of Fig. \ref{lenghtABO3} (d), confirm those deduced
from geometrical arguments in Section
\ref{microscopicdescription}.\\

\subsubsection{Anomalous character of the $c$-axis?}
\label{caxis}

When $T$ is reduced, the decrease of $c$ is slowing down as the
ferromagnetic transition at $T_C$ is approached, which ends by an
upturn below $T_C$, where $c$ increases with decreasing $T$ (see
Fig. \ref{thermal_expansion} (a)). The behavior of the $c$-axis
contrasts with those of the $a$- and $b$-axes, whose variations
are monotonic for $4<T<300$ K and are amplified below $T_C$ (see
Fig. \ref{thermal_expansion} (a)). The distortion of the ($a$,$b$)
plane results mainly from the GdFeO$_3$-type distortion, whose
modifications also control the uniaxial pressure dependences of
$T_C$ and $T_N$ for $\mathbf{p}\parallel\mathbf{a},\mathbf{b}$
(see Section \ref{GdFeO3}). In Sections
\ref{microscopicdescription} and \ref{latticeparam}, $c$ was shown
to be more sensitive than $a$ and $b$ to the elongations of the
TiO$_6$ octahedra, possibly related to a Jahn-Teller effect. The
sensitivity of $c$ to the distortions of the octahedra may be
related to the anomalous uniaxial pressure-dependences of $T_C$
and $T_N$ for $\mathbf{p}\parallel\mathbf{c}$ (see Section
\ref{GdFeO3}), but also to the anomalous behavior of the c-axis in
the spectral weight transfers of the optical conductivity
\cite{kovaleva07}. Our findings are in apparent contradiction to
theories according to which the Jahn-Teller distortion is an
essential prerequisite of ferromagnetism in YTiO$_3$ (Refs.
\onlinecite{mochizuki04,pavarini05}). Rather, the properties of
YTiO$_3$ seem to be reminiscent of those of
La$_{7/8}$Sr$_{1/8}$MnO$_3$, where a Jahn-Teller distortion is
fully suppressed at the onset of ferromagnetism \cite{Geck04}.

\subsection{High-temperature extrapolation} \label{extrapolation}

\begin{figure}[h]
    \centering
    \epsfig{file=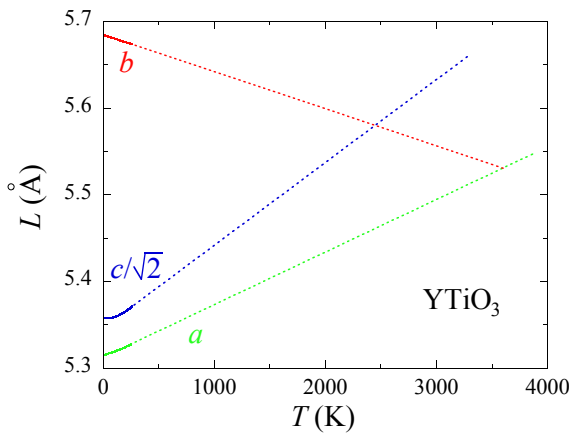,width=75mm}
    \caption{High temperature linear extrapolation of the lattice parameters
    $a$, $b$, and $c/\sqrt{2}$ of YTiO$_3$.}
    \label{lenght}
\end{figure}

Equivalently to uniaxial pressures and variations of the A$^{3+}$
ion size, increasing the temperature leads to a reduction of the
distortion in the ABO$_3$ perovskites. Indeed, the strong
anisotropy of the thermal expansion coefficients $\alpha_i$, shown
in Fig. \ref{thermal_expansion} (b), is governed by the effects of
temperature on the crystal distortion. Assuming constant thermal
expansion coefficients above room temperature, Fig. \ref{lenght}
shows high-temperature extrapolations, up to 4000 K, of the
lattice parameters $a$, $b$, and $c/\sqrt{2}$
\cite{note_lattice_param}. This plot indicates that, in YTiO$_3$,
a cubic structure with $a=b=c/\sqrt{2}$ \cite{notelatticecubic}
might be recovered around 3000-4000 K. However, this temperature
scale, characteristic of the lattice distortion, is inaccessible
since it is far above the melting temperature of YTiO$_3$.

\section{Conclusion}

The thermodynamic study of the perovskite system YTiO$_3$
presented here allowed us to extract information about the
ferromagnetic ordering and its coupling to the lattice
distortions. While the low-temperature specific-heat data, as well
as an Arrott plot of the magnetization close to $T_C$, are
consistent with a 3D Heisenberg picture of ferromagnetism,
deviations from this simple picture were observed in the thermal
expansion data at low temperature, where an unexpected power law
is found. Above $T_C$, a magnetic signal persists up to the
remarkably high temperature of $ 5 \times T_C$. Further work is
required to show whether models incorporating combined
spin-orbital fluctuations, instead of spin-only fluctuations,
could quantitatively describe this extended fluctuation regime.

Ehrenfest and Maxwell relations enabled us to extract the uniaxial
pressure dependences of the Curie temperature $T_C$ and of the
spontaneous moment $M_s$, which indicates that ferromagnetism is
stabilized by uniaxial pressures $\mathbf{p}\parallel\mathbf{a}$
and is destabilized by uniaxial pressures
$\mathbf{p}\parallel\mathbf{b},\mathbf{c}$ and by hydrostatic
pressure. We interpreted the uniaxial pressure dependences of
$T_C$ and $M_s$ obtained for
$\mathbf{p}\parallel\mathbf{a},\mathbf{b}$ as resulting from
uniaxial pressure-induced modifications of the GdFeO$_3$-type
distortion. A high sensitivity of the $c$-axis to an additional
distortion of the TiO$_6$ octahedra, possibly related to a
Jahn-Teller effect, is believed to be responsible for the
anomalous uniaxial pressure dependences of $T_C$ and $M_s$
observed for $\mathbf{p}\parallel\mathbf{c}$. This confirms that
both kinds of distortion play an important role for the formation
of ferromagnetism in YTiO$_3$, Jahn-Teller distortion being not a
necessary condition for ferromagnetism in YTiO$_3$. While the $a$-
and $b$-axes are more sensitive to the GdFeO$_3$-type distortion,
the $c$-axis is more sensitive to the elongations of the
octahedra. Finally, a high-temperature extrapolation of the
lattice parameters led to the onset of the distortion at a virtual
temperature of about 3000-4000 K. These results might be
considered to further develop models for the electronic properties
of the titanates.

\section*{Acknowledgments}

We acknowledge useful discussions with T. Schwarz, D. Fuchs, M.
Merz, R. Eder, O. Andersen, E. Pavarini, and G. Khaliullin. We
thank K. Fischer for synthesizing the EuS crystal studied here.
This work was supported by the Helmholtz-Gemeinschaft through the
Virtual Institute of Research on Quantum Phase Transitions and
Project VH-NG-016.


\begin{thebibliography}{20}
\bibitem{notespinwave} Well below the ordering temperature, ungapped ferromagnetic spin waves should
lead to a $T^d$ contribution to the specific heat and thermal
expansion, while ungapped antiferromagnetic spin waves should lead
to a $T^{d/2}$ contribution ($d$ is the dimensionality of
exchange). If there were the two kinds of branches in the
low-energy excitations (strong ferromagnetic and weak
antiferromagnetic-like spin waves), then two kinds of
contributions should be present in both the specific heat and
thermal expansion. In this case, the ferromagnetic contribution
should dominate the specific heat because it has a higher
intensity. Since the thermal expansion is related to the pressure
dependence of the entropy, it is dominated by the most
pressure-dependent energy scales. This means that weak
antiferromagnetic-like spin waves could dominate the
low-temperature thermal expansion if they were associated with a
strongly pressure-dependent energy scale.
\bibitem{notecriticalexponents} In a limit with no domain effect and with $H\rightarrow0$,
Equation (\ref{arrottequation}) implies that
$M=\left(T_C-T\right)^{\beta}$ for $T<T_C$ and
$\chi=\left(T-T_C\right)^{-\gamma}$ for $T>T_C$.
\bibitem{notecubicalpha} For symmetry reasons, the thermal
expansion coefficient $\alpha$ of a cubic system does not depend
on the direction of the measured length $L$ from which it is
derived.
\bibitem{notelatticecubic} In an undistorted perovskite structure, the
lattice parameters $a$, $b$, and $c$ of the orthorhombic unit cell
are related to the lattice parameter $a_0$ of the cubic unit cell
by $a=b=c/\sqrt{2}=a_0\sqrt{2}$.
\bibitem{note_jahn_teller} In YTiO$_3$, elongations by 3 \% of the octahedra were explained by a
Jahn-Teller effect in Ref.
\onlinecite{mochizuki04,pavarini05,akimitsu01,iga04}. In the
non-Jahn-Teller LaFeO$_3$ and YFeO$_3$, but also in LaTiO$_3$,
elongations, by about 1 \%, were reported in Ref.
\onlinecite{komarek07,duboulay95,zhou08} and are presumably
consequences from the GdFeO$_3$-type distortion.
\bibitem{note_lattice_param} $i=i_0(1+\Delta
L_i/L_i)$, where $i=a,b,c$ (cf. Fig. \ref{thermal_expansion} (a)),
and where $a_0=5.331$, $b_0=5.672$, and $c_0=7.602$ $\rm{\AA}$
were extracted from diffraction measurements at room temperature
\cite{kovaleva07}.
\end{thebibliography}
\end{document}